\documentclass[aps,prl,twocolumn,groupedaddress,showpacs]{revtex4}
\usepackage{epsfig}

\def\8{\infty}
\def\oh{\frac{1}{2}}

\def\undertext#1{\vtop{\hbox{#1}\kern 1pt \hrule}}

\def\dbyd#1#2{\frac{d#1}{d#2}}

\def\be{\begin{equation}}
\def\ee{\end{equation}}
\def\bea{\begin{eqnarray} & &}
\def\eea{\end{eqnarray}}

\def\rf#1{(\ref{#1})}

\def\rf#1{(\ref{#1})}

\def\rfs#1{Eq.~\rf{#1}}

\begin{document}


\title{Global large time dynamics and the generalized Gibbs ensemble}


\author{V. Gurarie}
\affiliation{Department of Physics, University of Colorado,
Boulder CO 80309}


\date{\today}

\begin{abstract}
We study the large time dynamics of a macroscopically large quantum systems under  a sudden quench. We show that, first of all, for a generic system in the thermodynamic limit the Gibbs distribution correctly captures the large time dynamics of its global observables. In contrast, for an integrable system, the generalized Gibbs ensemble captures its global large time dynamics only if the system can be thought of as a number of noninteracting uncorrelated fermionic degrees of freedom.  The conditions for the generalized Gibbs ensemble to capture the large time dynamics of local quantities are likely to be far less restrictive, but this question is not systematically addressed here. 
\end{abstract}

\maketitle
Recently a problem of an evolution of a quantum system after its Hamiltonian suddenly changed (a problem of quantum quench) attracted a lot of attention. A particular question which arises in this context is whether long time asymptotics of the observables of this system can be thought of as obeying a thermal Gibbs-like distribution 
\cite{Rigol2007}. The problem can be formulated in the following way. Suppose a system is initially in a quantum state $\left|  \psi \right>$ which is not an eigenstate of its Hamiltonian (but rather an eigenstate of the Hamiltonian before the quench). That state can be decomposed as $\left| \psi \right> = \sum_n c_n \left| n \right>$, where $\left| n \right>$ are eigenstates of the Hamiltonian after the quench. Then the time evolution of an expectation value of an observable can be found directly as
\be \left< \psi \right| {\cal O} (t) \left| \psi \right> = \sum_{nm} c_n^* c_m e^{i \left( E_n-E_m \right)t } \left< n \right| {\cal O} \left| m \right>. \ee
At long times the oscillating terms in this expression are supposed to die out (at least average away if one resorts to averaging over time, to avoid dealing with situations \cite{Spivak2004} where
oscillations may persist forever), leading to the time independent average
\begin{eqnarray} \left< \psi \right| {\cal O} (t) \left| \psi \right>_{t \ {\rm large}} &\rightarrow & \sum_{n} \rho_n \left< n \right| {\cal O} \left| n \right>, \ \rho_n =\left| c_n\right|^2, \cr \sum_n \rho_n&=&1.\end{eqnarray}

A question arises whether the averages over arbitrary probabilities $\rho_n$ are equivalent to averaging over a Gibbs ensemble, with an appropriately tuned temperature. Moreover, in some cases the system under consideration is integrable, that is, it has a large number of conserved quantities equal to half of the number of its degrees of freedom. Then the question is whether the averages over the arbitrary $\rho_n$ are equivalent to averaging over generalized Gibbs ensemble which includes all the integrals of motion in addition to energy in its construction \cite{Rigol2007,Imambekov2012,Tsvelik2012,Caux2012}. 

Before any general discussion, one must emphasize that if ${\cal O}$ is a local operator (in space), and if the integrals of motion of the system are local (which implies that they scale with the volume of the system), then there appears to be little doubt that any average of this operator in the thermodynamic limit (the limit of an infinite volume at fixed density) can be described by a (generalized) Gibbs distribution. To see that, one can follow any standard arguments regarding the behavior of a small subsystem interacting with a large bath whose role is played by the rest of the system. More detailed arguments specifically for the quenched transverse field Ising model are given here \cite{Essler2011,Essler2012a,Essler2012b}. Yet, suppose we are interested in global observables. Then it is not  obvious that the averages are given by the (generalized) Gibbs distribution. In particular, one may want to calculate the so-called diagonal entropy \cite{Polkovnikov2011}
\be  \label{eq:diag}  S_d = -\sum_n \rho_n \ln \rho_n.
\ee
The question may arise whether this entropy is equal to the entropy calculated using the (generalized) Gibbs ensemble. 

Here we show averaging a nonintegrable system over some arbitrary probability distribution $\rho_n$ in the thermodynamic limit is indeed equivalent to averaging over a Gibbs ensemble. However, averaging an integrable system over some arbitrary probability distribution is equivalent to averaging over generalized Gibbs ensemble only if the system can be though of having fermion-like degrees of freedom (which appears to be true for a large number of integrable systems), and when these degrees of freedom are not correlated. We will see, in particular, that quenches in the transverse field Ising model are not described globally by the generalized Gibbs ensemble because its effective degrees of freedom, while being fermionic, are correlated (while locally the quenches in the transverse field Ising model are indeed described by the generalized Gibbs ensemble \cite{Essler2011}). 

The reason why uncorrelated fermionic degrees of freedom are well described by a Gibbs-like ensemble is very simple. A single noninteracting fermionic degree of freedom can be thought of as a two level system, one level representing an empty fermionic state and one level representing a filled fermionic state. An arbitrary probability distribution of such a two level system is completely fixed by its average particle number. Therefore, if the distribution produces correct average particle number, it is an exact correct distribution. In particular, a Gibbs distribution for such a system is exact as well. This generalizes to a system consisting of a number of uncorrelated fermionic degrees of freedom. Conversely, we will see below that if a system cannot be thought as splitting into uncorrelated fermionic degrees of freedom, it appears unlikely that it can be described by its generalized Gibbs ensemble.

In what follows we first go over the derivation of the equivalence between the microcanonical and Gibbs ensembles for large closed systems, then discuss the diagonal entropy, the
generalized Gibbs ensemble, and finally go over the example of a quench in the transverse field Ising model whose large time behavior does not reduce to the generalized Gibbs ensemble. 

{\sl Gibbs ensemble.} Consider a large system with energy levels $E_n$. Its microcanonical entropy is defined by
\be \label{eq:micro} e^{S_m(E)} = \sum_n \delta(E - E_n) \, \delta E,
\ee
where $\delta E$ is a fixed energy interval. Let us show that this definition of the entropy coincides with the canonical entropy in the thermodynamic limit. Indeed,
\be \label{eq:ent} \frac{e^{S_m(E)}}{ \delta E} =\sum_n  \int_{-i\infty}^{i \infty} \frac{d\beta}{2\pi}  e^{ \beta \left(E - E_n \right) }   = \int_{-i \infty}^{i \infty} \frac{d\beta}{2\pi} e^{\beta \left( E - F(\beta) \right)},
\ee
where $F(\beta)$ is the canonical free energy
\be  F(\beta) =-\beta^{-1} \ln \left[ \sum_n e^{-\beta E_n} \right].
\ee
Free energy is proportional to the volume of the system, and so is $E$. Therefore, the integral over $\beta$ can be taken using the saddle point approximation. The saddle point equation reads
\be \label{eq:saddle} E- F - \beta \dbyd{F}{\beta}=0.
\ee
The canonical entropy is given by \be S(\beta) = \beta^2 \dbyd{F}{\beta}, \ee therefore \rfs{eq:saddle} really states that $F = E-S/\beta$. The solution of that equation $\beta(E)$ is the inverse temperature which corresponds to a specific energy $E$. Once this solution is known, the solution to the integral \rfs{eq:ent} is, within the saddle point approximation, $\exp \left[ {\beta \left( E- F(\beta) \right)} \right]$,
so it gives for the microcanonical entropy
\be \label{eq:microgibbs} S_m(E) = \beta \left( E- F(\beta) \right) = S(\beta).
\ee
This concludes the proof that these entropies are the same (more details can be found in Ref.~\cite{Gurarie2007aa}). The main criterion of the applicability of the proof is that the system is large so the quantity in the exponential in
\rfs{eq:ent} is proportional to its volume. Note that the energy interval $\delta E$ drops out in the thermodynamic limit and so will be omitted in further discussions. 

{\sl Diagonal entropy.} Let us show that quite generally, the diagonal entropy \rfs{eq:diag} is equal to the canonical entropy defined above. Indeed, by virtue of \rfs{eq:micro}, \rfs{eq:diag} is equal to 
\be \label{eq:diagent} S_d = - \int dE \, e^{S_m(E) + \ln \rho(E)} \ln \rho(E)
\ee
(here $\rho(E_n) = \rho_n$),
while the normalization condition for $\rho_n$ is \be \label{eq:norm} \int dE \, e^{S_m(E)+ \ln \rho(E)} = 1.
\ee
Entropy is an extensive quantity which scales with the volume. $\ln \rho(E)$ typically is also proportional to the volume (although this condition may be violated in some
specially constructed $\rho_n$). If so, we can take the integral over $E$ in both \rfs{eq:diagent} and \rfs{eq:norm} by the saddle point approximation. The saddle point occurs when
\be \label{eq:saddlenergy} \dbyd{S_m(E)}{E} + \frac 1 \rho \dbyd{\rho(E)}{E} =0.
\ee
 \rfs{eq:norm} gives, at the value $E$ corresponding to the saddle point,
\be \rho(E) = e^{-S_m(E)}.
\ee
Finally, the equation \rfs{eq:diagent} gives, in the same saddle point approximation, when combined with \rfs{eq:microgibbs},
\be \label{eq:theorem} S_d = S_m(E) = S(\beta).
\ee
This concludes the proof the the diagonal entropy and the Gibbs entropy are equal in the thermodynamic limit, just as observed in Ref.~\cite{Rigol2011} in their study of a time evolution of a non-integrable system. Let us reiterate  that this result is obtained assuming that, (a) the logarithm of the probabilities $\ln \rho_n$ are proportional to the volume of the system, which is natural (since the total density of states typically grows exponentially with the volume) but could be violated in some specific cases depending on the initial state $\left| \psi \right>$, (b) $\rho_n$ are smooth functions of energy, which again could be violated in some cases and (c) the saddle point condition \rfs{eq:saddlenergy} has only one solution, which again may be violated for some specially chosen $\rho_n$. 

{\sl Generalized Gibbs ensemble.} It is straightforward to generalize the above reasoning to the case when there are extra conserved quantities, such as the particle number
$N$. The 
definition of the microcanonical entropy, replacing \rfs{eq:micro}, is
\be e^{S_m(E,N)} = \sum_n \delta(E-E_n) \delta(N-N_n) \, \delta E,
\ee where $N_n$ is the number of particles in a given energy level. 
The rest of the formalism changes accordingly, with $F$ becoming the grand canonical free energy, the integration in \rfs{eq:ent} being over $\beta$ and the chemical potential $\mu$, and the integral in \rfs{eq:diagent} getting replaced by the integrals over $E$ and $N$. The end result is the same, that is, the diagonal and the Gibbs entropy are equal to each other in the thermodynamic limit.

However, integrable systems have a large number of conserved quantities, equal to half of the number of their degrees of freedom. Consider for example a Tonks gas which maps into a system of  one dimensional noninteracting fermions. Its conserved quantities are the fermionic occupation numbers $I_k$, where $k$ labels one dimensional momenta ($I_k$ are real numbers between $0$ and $1$ while we reserve the notation $n_k$ for an integer taking values $0$ and $1$). Its microcanonical entropy is given by
\begin{eqnarray} \label{eq:gge} e^{S_m(I)} &=& \sum_{n_k=0,1} \prod_k \delta(I_k-n_k) \cr  &=& \sum_{n_k=0,1} \int \left[ \prod_k \frac{d\mu_k}{2\pi} \right]  e^{\sum_k \mu_k \left( n_k-I_k \right)},
\end{eqnarray}
where $\mu_k$ are the generalized chemical potentials. Doing the sum over $n_k$ gives
\be \label{eq:gge1} e^{S_m(I)} = \int \left[ \prod_k \frac{d\mu_k}{2\pi} \right] e^{-\sum_k \left( \mu_k I_k +\Omega(\mu_k) \right)}.
\ee
Here \be
\Omega(\mu) = - \ln \sum_{n=0,1} e^{\mu n} = - \ln \left[ 1+ e^{\mu} \right],
\ee
the generalized free energies. The Legendre transform of $\Omega$ defines the generalized Gibbs entropy by
\be \label{eq:entropygge} S(I) =- \sum_k \left(  \mu_k I_k +\Omega(\mu_k) \right), \ I_k =- \dbyd{\Omega(\mu_k)}{\mu_k}.
\ee

As before, the probabilities to occupy various states in this system are given by $\rho(n)$ where $n$ stands for a collection of $n_k$,  each $n_k$ for each $k$ takes two values, $0$ or $1$. Now suppose $\rho(n)$ factorizes into a product \be \rho(n) = \prod_k \rho_k({n_k}),\ee that is, each degree of freedom is independent random variable in the initial state. 
The normalization condition states that 
\be \label{eq:norm1} \rho_k(0)+\rho_k(1)=1
\ee for each $k$. Then the 
diagonal entropy is given by the sum of the diagonal entropies
\be \label{eq:correct} S_d = - \sum_k \left[ \rho_k(1) \ln \rho_k(1) + (1-\rho_k(1)) \ln (1-\rho_k(1)) \right].
\ee
One can establish by inspection that this entropy is exactly equal to the Gibbs entropy defined in \rfs{eq:entropygge}, if one identifies
\be \label{eq:probid} I_k = \sum_{n_k=0,1} n_k \rho_k(n_k) = \rho_k(1),
\ee
that is, $I_k$ is the average occupation number of the initial state. Indeed, 
we write
\be I_k = -\dbyd{\Omega(\mu_k)}{\mu_k} = \frac{1}{e^{-\mu_k}+1}.
\ee
This allows us to express $\mu_k$ in terms of $I_k$ and evaluate
\begin{eqnarray} \label{eq:entgibbs} S &=& - \sum_k \left(  \mu_k I_k + \Omega (\mu_k)  \right)  = \cr &&  - \sum_k \left( I_k \ln I_k + (1-I_k) \ln (1-I_k) \right),
\end{eqnarray} which of course exactly coincides with the answer for $S_d$.

Moreover, the probabilities of observing a state with an occupation $n_k$ in the Gibbs ensemble are given by
\be \label{eq:gibbsprobdef} \rho_{k}^G(n) = e^{\Omega(\mu_k)+n_k \mu_k}.
\ee
It is easy to see that these probabilities give
\be \label{eq:gibbsprob} \rho_{k}^G(0) = 1-I_k, \ \rho_k^G(1) = I_k,
\ee that is, precisely the right values of the probabilities $\rho_k$ as given in \rfs{eq:probid}. That is, the Gibbs ensemble reproduces the distribution $\rho(n)$ exactly, not just for the diagonal entropy, but also for all the averages.

On the one hand, this appears to be nothing but the manifestation of the theorem \rfs{eq:theorem}. On the other hand, the conditions under which the theorem \rfs{eq:theorem} was proven no longer hold true. Indeed, by introducing a large number of conserved quantities, whose number scales with the system size, we have a situation where each individual conserved quantity does not scale with the system size and the saddle point approximation cannot be valid. Let us look at it in more details. 

\rfs{eq:gge1} allows us to rewrite the diagonal entropy as
\be \label{eq:def1} S_d = - \sum_k \int  \frac{d\mu dI}{2\pi}  \ln \rho_k(I) \,e^{\ln \rho_k(I)- \mu I -\Omega  },
\ee
while the normalization condition  \rfs{eq:norm1} reads
\be  \label{eq:def2}  \int  \frac{d\mu dI}{2\pi}  \,e^{\ln \rho_k(I)- \mu I -\Omega  }=1
\ee
for each $k$. Eqs.~\rf{eq:def1} and \rf{eq:def2} constitute the exact definition of $S_d$. 
There is no reason why saddle point approximation is valid for the integrals in these equations, since the expressions in the exponential are not large in the limit of large system size; in fact, they do not depend on the size of the system. Nevertheless, computing these integrals via the saddle point approximation we find the correct answer \rfs{eq:correct}, the same as would have been found had these integrals been computed exactly. At the same time, the saddle point approximation gives for the value of the integral
\rfs{eq:def1} precisely $S$, the generalized Gibbs entropy defined in \rfs{eq:entropygge}. This appears to be the  the origin why the Generalized Gibbs ensemble well describes the time evolution of integrable systems. 

To see how this works out, we note that evaluating the integral \rfs{eq:def2} via a saddle point approximation gives
\be \rho_k(I) = e^{\Omega + \mu I}, \ I = -\dbyd{\Omega}{\mu}.
\ee
This means that \be \rho_k(I) = e^{-S(I)},\ee where $S$ is the entropy calculated within the generalized Gibbs ensemble. Substituting this into \rfs{eq:def1}, we find that $S_d$, evaluated via a saddle point approximation, coincides with $S$.

While it is not clear why the saddle point approximation gives exact answers in this case, it appears to be crucial that we are describing a system which consists of a number of independent degrees of freedom, each of them taking only two values. If any of these conditions are no longer true, the Gibbs ensemble stops describing our system. Indeed, the arguments leading to \rfs{eq:gibbsprob} already hint that it was crucial that we were dealing with a degree of freedom which takes only two values. For such a variable, if you know its average $I$ you also know the probabilities that it takes these two values, which must be the origin why the Gibbs distribution is equivalent to an arbitrary distribution in this case. 

For example,  consider a situation where the initial state has $\rho(n_k)$ which correlate some of the $n_k$ \cite{Pustilnik2008}. For example, suppose $\rho(n_1, n_2)$ does not split into a product of two terms, depending on $n_1$ and $n_2$ respectively. Then the contribution of these two $n$ to the diagonal entropy becomes 
\begin{eqnarray} && S_d =  - \int  \frac{d\mu_1 d\mu_2 dI_1 dI_2}{(2\pi)^2}  \ln \rho(I_1,I_2) \times \cr && e^{\ln \rho(I_1,I_2)- \mu_1 I_1- \mu_2 I_2 -\Omega(\mu_1)-\Omega(\mu_2)  }.
\end{eqnarray}
There is no reason why the saddle point approximation when applied to this integral should give the same answer as if this integral is calculated exactly. And indeed, a straightforward evaluation of the saddle point approximation shows that the Gibbs entropy and the diagonal entropy are not equal. On the other hand, If $\rho(I_1,I_2) = \rho_1(I_1) \rho_2(I_2)$, then the saddle point approximation gives the correct answer, as this reduces to the example of uncorrelated degrees of freedom considered previously.

Moreover, the Gibbs entropy \rfs{eq:entropygge} in these example of two correlated degrees of freedom by construction depends on two parameters $I_1$ and $I_2$, each being the average of $n_1$ and $n_2$ respectively, with respect to $\rho$. Yet the exact entropy depends on three parameters, the entries of the matrix $\rho(n_1, n_2)$ constrained by the normalization condition. It is possible to change $\rho$ so that $S_d$ changes yet $I_1$, $I_2$, and thus the Gibbs entropy $S$, as well as Gibbs probabilities $\rho^G$ defined in \rfs{eq:gibbsprobdef}, do not change. Thus we see quite generally that $S_d$ and $S$ cannot be equal to each other barring
some coincidence. 

We deduce from here that the generalized Gibbs ensemble correctly describes those systems which can be mapped into independent uncorrelated fermion-like degrees of freedom. Introducing any kinds of correlations results in corrections to the generalized Gibbs ensemble. Those corrections may be small and may not be immediately seen in numerical simulations (and may explain the discrepancy between the generalized Gibbs and diagonal entropies in Ref.~\cite{Rigol2011}). 

{\sl A quench in the transverse field Ising model.} For an example of a quench in a free fermion-like problem with strong correlations, consider a quench in the transverse field Ising model, which has been a subject of intense studies recently  \cite{Cugliandolo2012}. The model is given by the Hamiltonian
$H = - \gamma \sum_j \tau^z_j - \beta \sum_j \tau^x _j \tau^{x}_{j+1}$.  
We consider a situation where the system described this Hamiltonian is in its ground state, and subsequently the parameters $\gamma$ and $\beta$ are changed abruptly. 
Under the Jordan-Wigner transformation $\tau^x_j+i \tau^y_j = 2 a^\dagger_j \exp\left(i \pi \sum_{\ell<j} a^\dagger_\ell a_\ell \right)$ this maps into a free fermion problem with the Bogoliubov-de-Gennes-like Hamiltonian
\be H=\sum_{k>0} h_k \left( a^\dagger_k a_k - a_{-k} a^\dagger_{-k} \right) + \Delta_k \left( a^\dagger_{k} a^\dagger_{-k} + a_{-k} a_{k} \right),\ee  where \be \label{eq:par} h_k = 2 (\gamma + \beta \cos k), \ \Delta_k = 2 \beta \sin k.\ee 
This Hamiltonian is diagonalized by a Bogoliubov transformation from $a_k$ to $b_k$ (known as Bogoliubov particles). 
The ground state is a vacuum of $b$-particles, equivalent to the BCS ground state. 

If the parameters of the Hamiltonian suddenly change the old ground state becomes a linear superposition of the eigenstates of the new Hamiltonian, which is diagonalized in terms of new ``after the quench" fermions $c_k$. It is easy to see that this state can be written as
\be \left| GS_{\rm old} \right> = \sum_{k>0} \left( u_k + v_k c^\dagger_{k} c^\dagger_{-k} \right) \left| 0 \right>,
\ee
where
\be u_k^2 = \oh \left( 1 + \frac{h_k^{0} h_k + \Delta_k^{0} \Delta_k }{\sqrt{  \left( \left(   \Delta_k^{0} \right)^2 + \left( h_k^{0} \right)^2 \right) \left( \Delta_k^2 +h_k^2 \right)} } \right),
\ee
$v_k^2=1-u_k^2$,
and $h^{0}_k$, $\Delta^{0}_k$ and $h_k$ and $\Delta_k$ are expressed in terms of the parameters of the problem before the quench and after the quench respectively, according to \rfs{eq:par}.

 In other words, there are strong correlations between particles of momentum $k$ and $-k$ which appear in pairs only. 
It is easy to check now that its diagonal entropy is equal to 
\be S_d = - \sum_{k>0} \left[ u_k^2 \ln u_k^2 + \left(1-u_k^2 \right) \ln \left( 1- u_k^2 \right) \right].
\ee
At the same time, the Gibbs entropy which can be computed in a straightforward way with the formalism introduced above in \rfs{eq:entgibbs} is
\be S = - 2 \sum_{k>0}\left[ u_k^2 \ln u_k^2 + \left( 1-u_k^2\right) \ln \left(1-u_k^2\right) \right],
\ee that is twice as large. It is not equal to the diagonal entropy since it does not take into account correlations between the free fermions.

The discussion so far was directly applicable to models which maps into free fermions. However it is well known that a wide variety of integrable systems behave as a collection of fermionic degrees of freedom. Typically one needs to specify a set of integers, and the values of the so-called rapidities fixing the wave function depend on whether a particular integer is present  in the set, leading to a fermion-like degree of freedom $n_k$ specifying whether a particular integer $k$ is present \cite{KorepinBook}. Thus everything discussed here seems to apply to quite general integrable systems. 

The author is grateful to the participants of the KITP program ``Quantum Dynamics in Far from Equilibrium Thermally Isolated Systems" and especially to A. Polkovnikov and F. Essler for discussions concerning the subject of this paper. The work described here is supported by the NSF grants no. PHY-1211914, DMR-1205303, and PHY-1125915.

\bibliography{references}

\end{document}